\documentclass[twocolumn,showpacs,amsmath,amssymb,floatfix]{revtex4}

\usepackage[dvips]{color}
\usepackage{graphicx}
\usepackage{pifont}
\usepackage{amstext}
\usepackage{amsmath}
\usepackage{graphics,amssymb}
\usepackage{boxedminipage}
\usepackage{bbm}

\begin{document}
%
%
\title{Spin gap and string order parameter in the ferromagnetic\\ Spiral
       Staircase Heisenberg Ladder: a quantum Monte Carlo study}
\author{C. Br\"unger, F. F. Assaad}
\affiliation{Institut f\"ur Theoretische Physik, Universit\"at W\"urzburg,
             D-97074 W\"urzburg, Germany}
\author{S. Capponi, F. Alet}
\affiliation{Laboratoire de Physique Th\'eorique, IRSAMC, Universit\'e
             Paul Sabatier, CNRS, 31062 Toulouse, France}
\author{D. N. Aristov}
\altaffiliation[On leave from] { Petersburg Nuclear Physics
Institute, Gatchina  188300, Russia} \affiliation{Institut f\"ur
Theorie der Kondensierten Materie,
             Universit\"at Karlsruhe, Germany}
\affiliation{Center for Functional Nanostructures, Universit\"at
             Karlsruhe, 76128 Karlsruhe, Germany}
\author{M. N. Kiselev}
\affiliation{The Abdus Salam International Centre for Theoretical
             Physics, Strada Costiera 11, Trieste, Italy}
\begin{abstract}
We consider a spin-1/2 ladder with a ferromagnetic rung coupling
$J_\bot$ and inequivalent chains. This model is obtained by a
twist ($\theta$) deformation of the ladder and interpolates
between the isotropic ladder ($\theta=0$) and the $SU(2)$
ferromagnetic Kondo necklace model ($\theta=\pi$). We show that
the ground  state in  the  ($\theta$, $J_\bot$) plane  has a  finite
string order parameter characterising  the Haldane phase. Twisting
the chain introduces a new energy scale, which we interpret in
terms of a Suhl-Nakamura interaction. As a consequence we observe
a crossover in the scaling of the spin gap at weak coupling  from
$\Delta/J_{\|}\propto J_\perp/J_\| $ for $\theta < \theta_c \simeq 8\pi/9$ to
$\Delta/J_{\|}\propto (J_\perp /J_{\|})^{2} $  for $\theta > \theta_c$. Those results are
obtained on the basis of large scale Quantum Monte Carlo calculations.
\end{abstract}
\maketitle
%
%
Low-dimensional quantum magnets are fascinating objects from both 
experimental and theoretical points of view. 
Spin-$1/2$ ladders have been
widely studied  and interpolate  between the physics
of one-dimensional antiferromagnetic (AF) spin chains and two-dimensional
systems~\cite{Dagotto96}. In the one-dimensional (1D) case, there is an important
mapping between spin-$1/2$ Heisenberg AF chains and Luttinger liquids
~\cite{Lieb61} which allows to treat such chains by means of exact
fermionization and bosonization methods, resulting in a well-understood gapless
phase~\cite{Affleck86}. Coupling identical chains to form a spin ladder is
however not a trivial task from a theoretical point of view
~\cite{Gogolin98,sch1}. Indeed, the coupling is a relevant perturbation and,
 up to logarithmic corrections,  opens a gap proportional to the interchain
coupling $J_\bot$ ~\cite{Shelton96,Larochelle04}.
%
\begin{figure}
\includegraphics[width=0.45\textwidth]{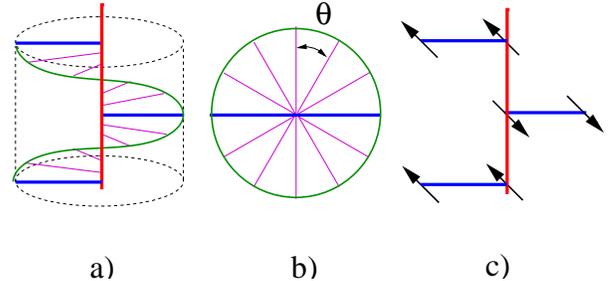}\\
\caption{\label{fig:f1a} (Color online) (a) Sketch of Spiral
Staircase Heisenberg Ladder.
         (b) View of the model from the top. (c) For $\theta=\pi$, the model
     maps to the 1D SU(2) ferromagnetic Kondo necklace model \cite{com1}.}
\end{figure}
%
In this paper, we will focus on the opening of the spin gap for the case of two
inequivalent chains coupled with a ferromagnetic rung coupling $J_\bot <0$.
This model is dubbed the Spiral Staircase Heisenberg Ladder:
%
\begin{eqnarray}
\hat{H}&=&J_\| \sum_i\left(
\hat{\mathbf{S}}_{1,i}\cdot \hat{\mathbf{S}}_{1,i+1}+
\cos^2\left(\theta/2\right)
\hat{\mathbf{S}}_{2,i}\cdot \hat{\mathbf{S}}_{2,i+1}\right)
\nonumber\\
&+&J_\perp\sum_{i} \hat{\mathbf{S}}_{1,i}\cdot \hat{\mathbf{S}}_{2,i}.
\label{hh1}
\end{eqnarray}
%
Here $\hat{\mathbf{S}}_{\alpha,i}$ is a spin-$1/2$ operator on leg
$\alpha$ and lattice site $i$. $J_\|>0$ sets the energy scale  and
the interchain coupling is taken to be ferromagnetic $J_\bot<0$.
Geometrically,  this model may be interpreted as a result of twist
deformation of a 2-leg ladder (Fig.~\ref{fig:f1a}a) with twist 
performed along one of the legs. Such a spiral structure is
characterized by the angle $\theta$ (see Fig.~\ref{fig:f1a}b) and
interpolates between the isotropic ladder ($\theta=0$) and a
ferromagnetic  SU(2) Kondo Necklace \cite{com1} model ($\theta =
\pi)$ \cite{Don77,kak05a,Kiselev05a,Aristov07}.  A motivation to
study this specific geometry comes from the fact that a realization 
of the model schematically presented in Fig.~\ref{fig:f1a}c was
synthesized as a stable organic biradical crystal PNNNO
\cite{Hosokoshi99}. Possible candidates for realizations with twist angle 
$0<\theta<\pi$ might be found in the families of
molecular chains decorated by magnetic radicals.

In the strong coupling limit, $|J_\bot/J_\parallel| \gg 1$, the
model maps onto the spin-1 Heisenberg chain with effective exchange
interaction $J_\mathrm{eff}=\tfrac{J_{\|}}{4}\left(1+\cos^{2}(\theta/2)\right)$. This phase
has a spin gap \cite{Haldane83} given by $\Delta_{H}/J_\mathrm{eff}=0.41048(6)$
~\cite{Todo01} and is characterized by a non-local string order parameter~\cite{Nijs89} 
(see a recent discussion in~\cite{Rosch07}):
%
\begin{equation}
\langle \hat{\mathcal O}_s (n)\rangle =\langle\hat{S}^{z}_{n_{0}}
\exp\left[i\pi\sum^{n_{0}+n}_{j=n_{0}} \hat{S}^{z}_{j}\right]
\hat{S}^{z}_{1,n+n_{0}}\rangle \label{String}
\end{equation}
%
with $\hat{S}^{z}_{j}=\hat{S}^{z}_{1,j}+\hat{S}^{z}_{2,j}$. The expectation value 
picks up the hidden antiferromagnetic
ordering. At weak couplings, the analysis depends on the twist angle $\theta$.
For small twist angles (i.e. close to the isotropic case), one can rely on the
bosonization and  numerical  results of Refs. \onlinecite{Shelton96,Larochelle04}
which yield a spin gap proportional to $|J_{\bot}|$ up to logarithmic
corrections. On the other hand, at $\theta = \pi$ the spin velocity on the
second leg vanishes thus inhibiting the very starting point of Ref.~\onlinecite{Shelton96}.  
Alternative approaches such as a mean-field theory based on a
Jordan Wigner transformation,  which yields the correct result for the
isotropic ladder, predicts a spin gap  $\Delta \propto J_{\bot}^2 /J_{\|} $
at $\theta = \pi$ \cite{Brunger07}.
A flow equation calculation has recently been carried out for the  $SU(2)$ Kondo
necklace model \cite{Essler07}, (i.e. $\theta = \pi$  in Eq. (\ref{hh1})) and is interpreted 
in terms of the onset of a spin gap irrespective of the value of $J_{\bot}/J_{\|}$.
%
%
%
To disentangle this situation, we have performed large scale quantum Monte
Carlo (QMC) simulations of the ferromagnetic spiral staircase model.
Two variants of the loop algorithm~\cite{Evertz03} were applied. For the string order parameter
and the spin-spin correlation functions, we used a discrete time algorithm and
extract the spectral functions via stochastic analytical
continuation schemes \cite{Sandvik98,Beach04}. For the spin gap calculation, a 
continuous time loop algorithm was used, where the gap is
calculated by a second moment estimator of the correlation length \cite{Todo01}.
%
\begin{figure}
\raggedright
\includegraphics[width=0.45\textwidth]{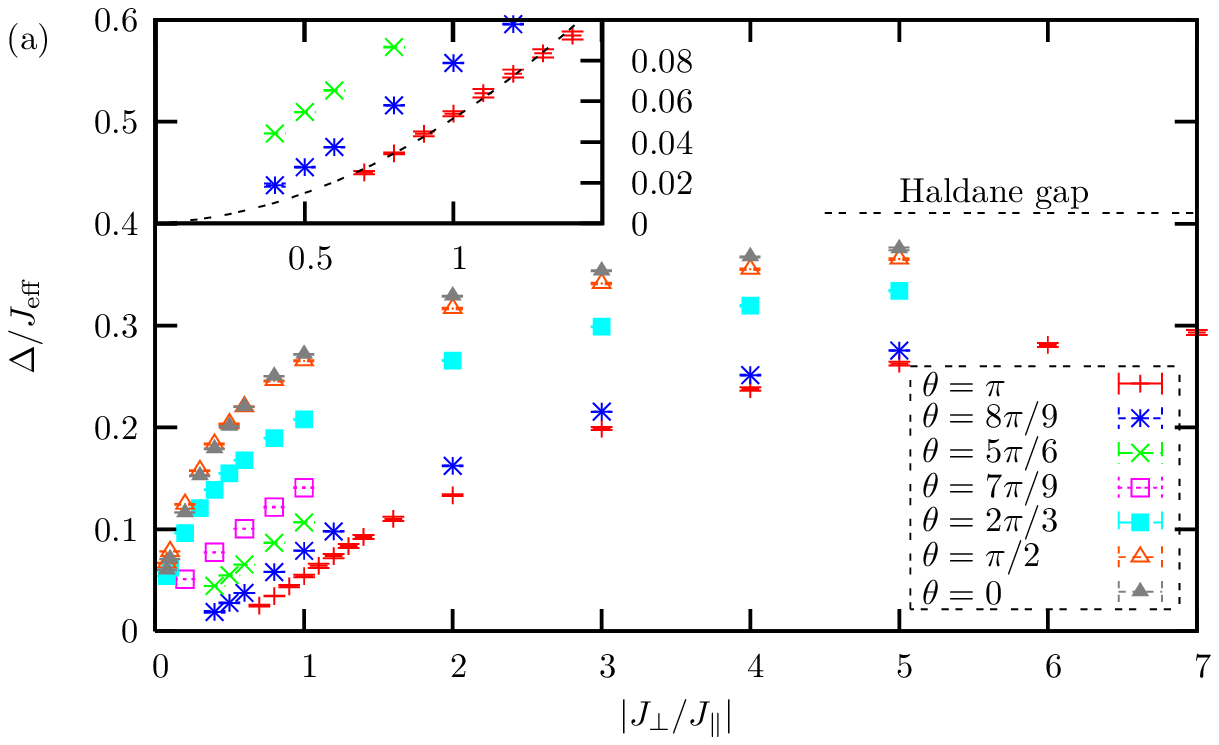} \\
\includegraphics[width=0.48\textwidth]{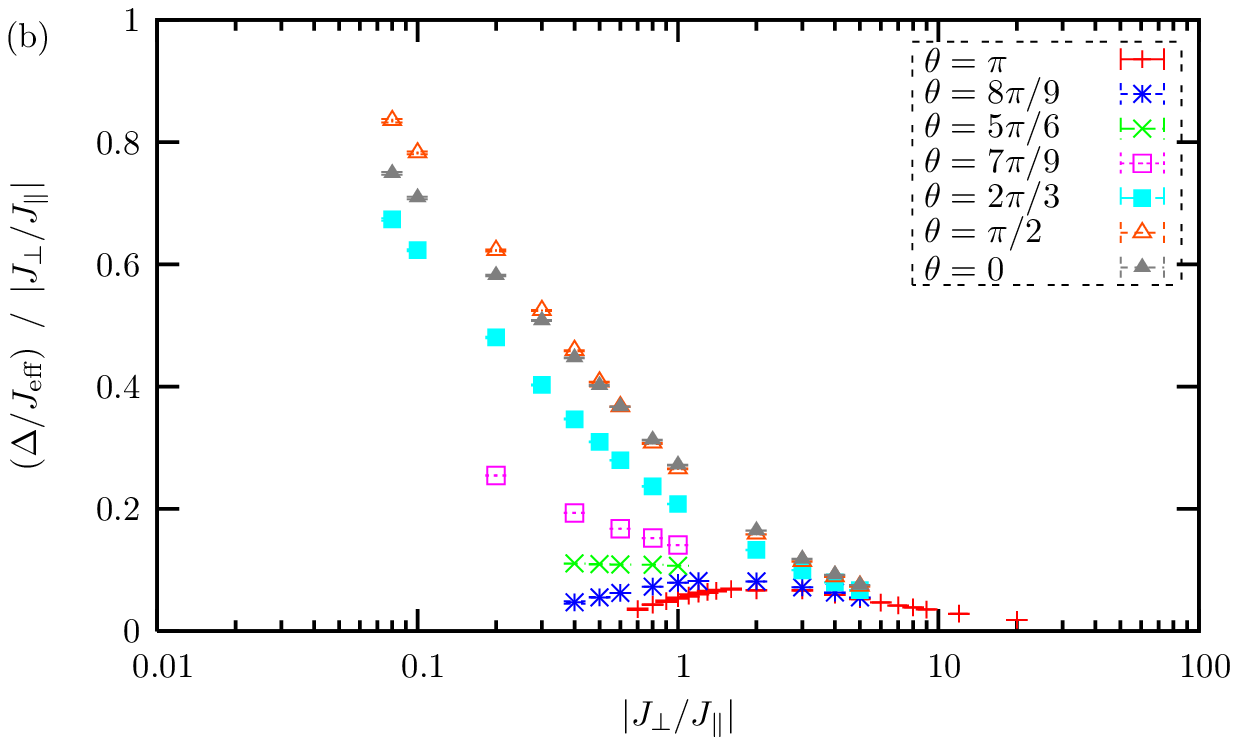}
\caption{\label{Gap.fig}(Color online) (a) Spin gap
$\Delta(J_{\bot})$ as a function of $|J_{\perp}/J_{\parallel}|$ for different twist
         angles $\theta$. The gap is rescaled by
     $J_\mathrm{eff}=\tfrac{J_{\|}}{4}\left(1+\cos^{2}(\theta/2)\right)$
         such that in the large-$|J_{\bot}|$-limit, it converges asymptotically
     toward the Haldane gap of a spin-$1$ chain. At weak couplings, 
         we have carried out QMC simulations up to $\beta J_{\|}=2500$ and
     $2\times 512$ spins to ensure size and temperature convergence.  
     Inset: zoom on the weak coupling region. 
     (b) Results for spin gap on a semi-logarithmic scale.}
\end{figure}
%

Our results for the spin gap in units of $J_\mathrm{eff}$ in the 
($\theta$,$J_{\bot}$)
plane are plotted in Fig. \ref{Gap.fig}. Enhancing the
twist angle from $\theta = 0 $ to $\theta = \pi/2$ leaves the spin
gap, measured in units of $J_\mathrm{eff} $, next to invariant thereby
showing that a {\it small} twist  is an irrelevant perturbation \cite{com4}.
For larger values of $\theta$, $\Delta$ is suppressed, and in the limit $\theta
= \pi$  the approach to the Haldane  value in the limit $J_{\bot}
\rightarrow -\infty$ is surprisingly  slow. At  small values of
$|J_{\perp}/J_{\parallel}|$, and $\theta = 0$  we reproduce the
results of Ref. \onlinecite{Larochelle04} namely  $\Delta \propto J_\bot$
(see Fig. \ref{Gap.fig}b). Here and in what follows, we 
neglect logarithmic corrections in our discussion. Fig.
\ref{Gap.fig}b shows that this weak coupling behavior of the spin
gap is sustained  up to $\theta < \theta_c \simeq 8 \pi /9 $. Beyond
this critical angle \cite{com2}, the data allows for different
interpretations. Let us concentrate on the twist angles $\theta =
8\pi/9$  and $\theta= \pi$. A linear extrapolation of the data would
lead to the vanishing of the spin gap at a finite critical value of
$J_\bot$. 
However, in this parameter range, we find a finite string order parameter (see below), 
incompatible with a gapless phase.  
As suggested by a Jordan-Wigner mean-field 
analysis \cite{Brunger07}, we instead  assume the existence of an
inflection point and fit the data to a quadratic form  in the limit
$J_\bot \rightarrow 0$  (see inset of Fig. \ref{Gap.fig}a). Let us note, however,
that we cannot exclude the possibility of an exponential scaling.
%
\begin{figure}
\centering
\begin{center}
\includegraphics[width=0.45\textwidth]{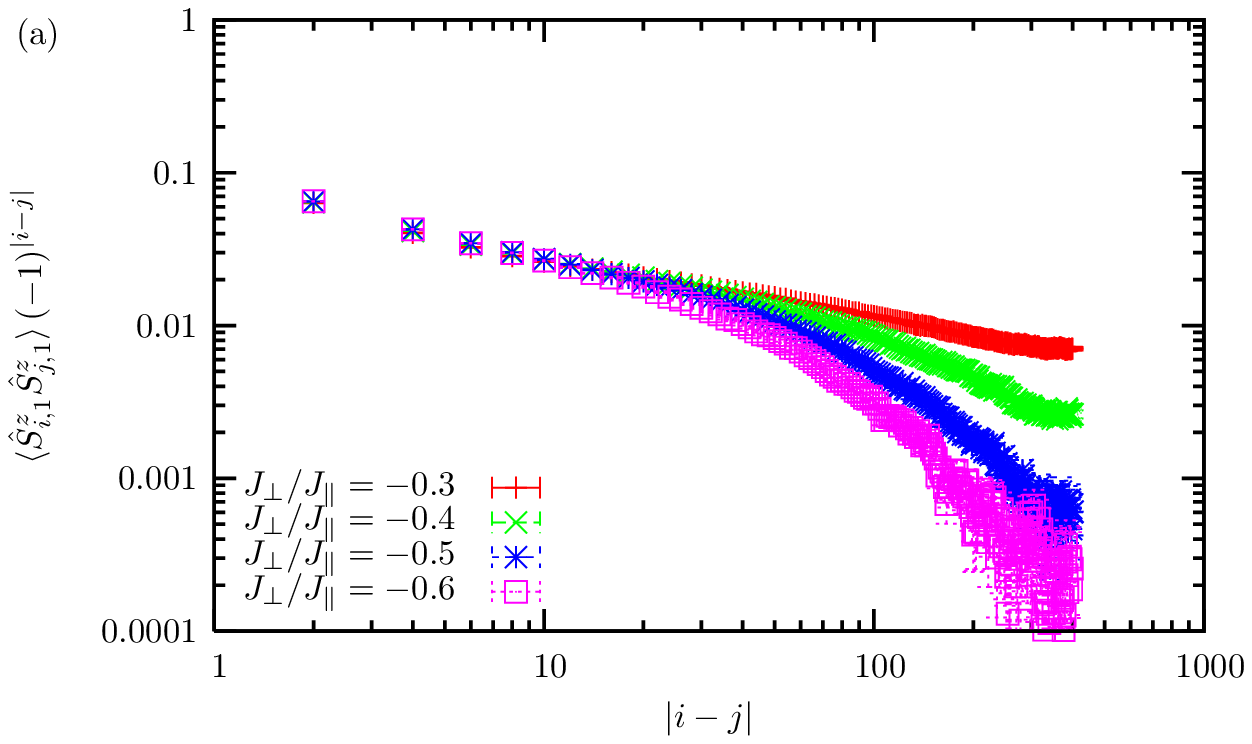}
\includegraphics[width=0.45\textwidth]{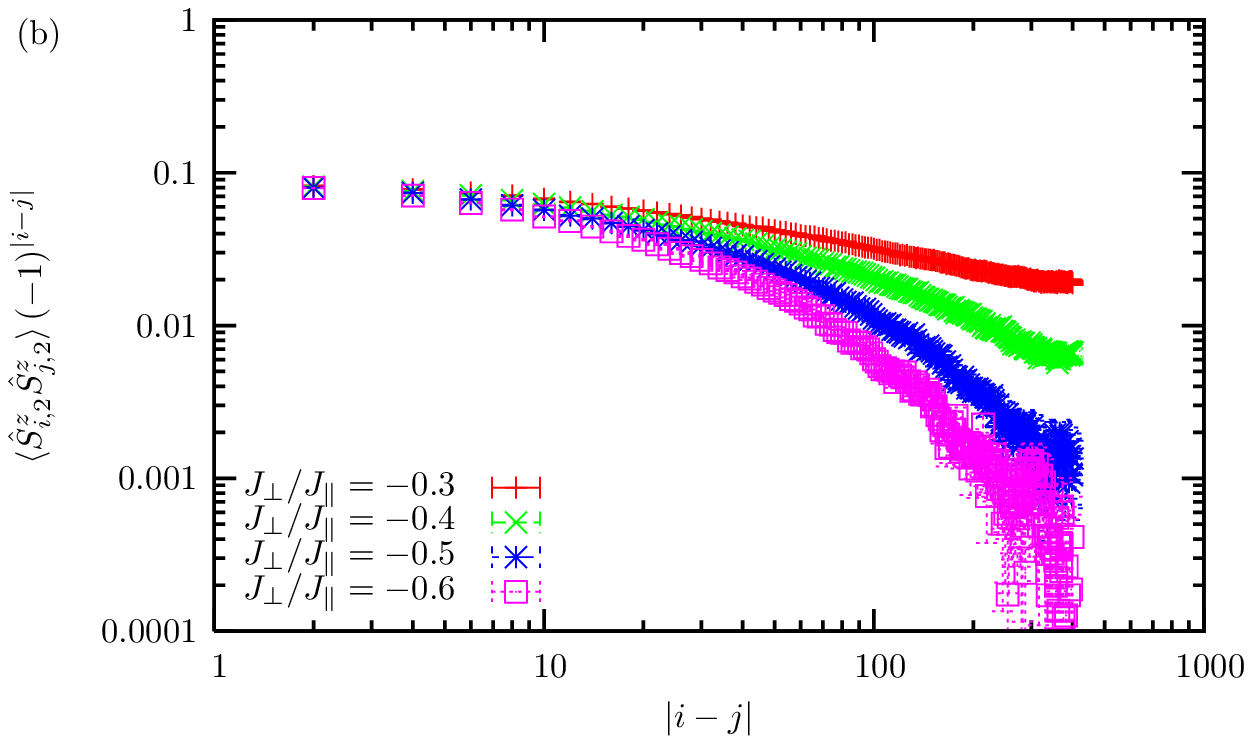}
\end{center}
\caption{\label{Spin.fig} (Color online) Spin-spin correlation
function on both legs for
         the Kondo necklace model ($\theta=\pi$) at different couplings
         $J_{\bot}/J_{\|}$ on a $2\times 800$ lattice. Simulations are carried
	 out at $\beta J_{\|}=7000$ ($J_{\bot}/J_{\|}=-0.3,-0.4$), 
	 $\beta J_{\|}=5000$ ($J_{\bot}/J_{\|}=-0.5$) and 
	 $\beta J_{\|}=2000$ ($J_{\bot}/J_{\|}=-0.6$).}
\end{figure}
%

The scaling of the spin gap at $\theta >\theta_c$ implies a rapid
increase of the spin correlation length $\xi \propto J_{\|}/\Delta$. For $\theta=\pi$ and 
$J_\perp/J_\| = -0.5 $, spin correlations decay exponentially with
characteristic length scale $\xi \simeq 115$ (see Fig.
\ref{Spin.fig}). At $J_\perp/J_\| = -0.3$ no sign of exponential
decrease is apparent on the considered $2 \times 800$ lattice. This
is consistent with a spin gap decreasing as $J_{\perp}^2/J_{\|}
$ (or quicker). Indeed, such as scaling  leads to  $\xi \geq 300 $ which is  comparable to the largest distance $L/2=400$ 
accessible in our simulation of a $2 \times 800$ lattice. 

On length scales $ |i-j| < \xi $ the spin-spin correlation functions follow  a 
slow power law.  In particular the data of Fig.~\ref{Spin.fig} 
at $J_\perp/J_\| = -0.3$ are consistent with
$S(|i-j|) \propto (-1)^{|i-j|} |i-j|^{-1/3} $.  
%
At $\theta = \pi$,
the effective interaction on the second leg is set by the 
Suhl-Nakamura (SN)~\cite{com3} interaction~\cite{Suhl58}. 
In second order perturbation
theory, without attempting any self-consistent calculation, this
interaction takes  the form in $ J_{SN} (q) \propto J_{\perp}^2
\chi_s(q,\omega=0) $ in Fourier space.  Here, $\chi_s(q,\omega=0) $
is the spin susceptibility of the spin $1/2$-chain. A first step
towards a self-consistent treatment is to allow for a gap, $\Delta$,
in $\chi_s(q,\omega=0)$. Thereby and in real space we expect SN
interaction to have a range set by $\xi$. We interpret the
above  mentioned very slow decay of the spin-spin correlations on both legs
and on a length scale  set by $\xi$ as a consequence of the SN
interaction.
%
%
\begin{figure}
\includegraphics[width=0.45\textwidth]{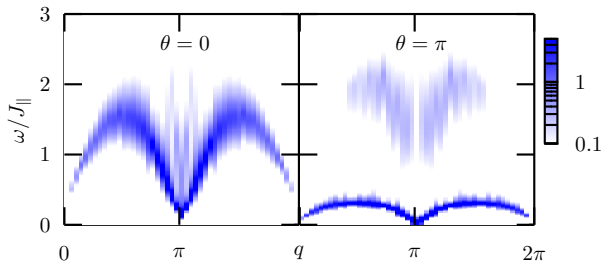}
\caption{(Color online) Dynamical spin-spin correlations at
         $J_\perp/J_\| = 1$ for the ladder
         system ($\theta=0$) and the Kondo necklace model ($\theta=\pi$).
         Here we consider a bonding combination of the spins across the rungs. 
	 ($\beta J_{\|}=200$, $L=100$)
     \label{SpinDyn.fig}}
\end{figure}
%
The SN interaction at $\theta = \pi$ sets a new low-energy scale
in the problem,  corresponding to the slow dynamics of the spins
degrees of freedom on the second leg.  Due to the ferromagnetic
coupling between the chains, this slow dynamics will equally
dominate the low energy  physics of the spins on the first chain.
This new energy scale is also apparent in the dynamical spin structure
factor $S(q,\omega)$ plotted in Fig. \ref{SpinDyn.fig}.  As
apparent, a narrow magnon band emerges as the angle $\theta$ grows
from $0$ to $\pi$. To lend support to the interpretation in terms
of the SN interaction, we have checked with exact diagonalization
methods that the width  of the magnon  band at $\theta = \pi$
indeed scales as $J_\perp^2/J_\|$ in the weak interleg coupling
limit (data not shown). 
In the vicinity of $\theta = \pi$, we hence expect that the low energy
effective model is given by a spin-1 Heisenberg chain with exchange coupling set by the SN
interaction. Assuming the validity of this low energy model, we
predict a spin gap which scales as $J_{SN} \propto
J_{\bot}^2/J_{\|} $.

\begin{figure}
\raggedright
\includegraphics[width=0.45\textwidth]{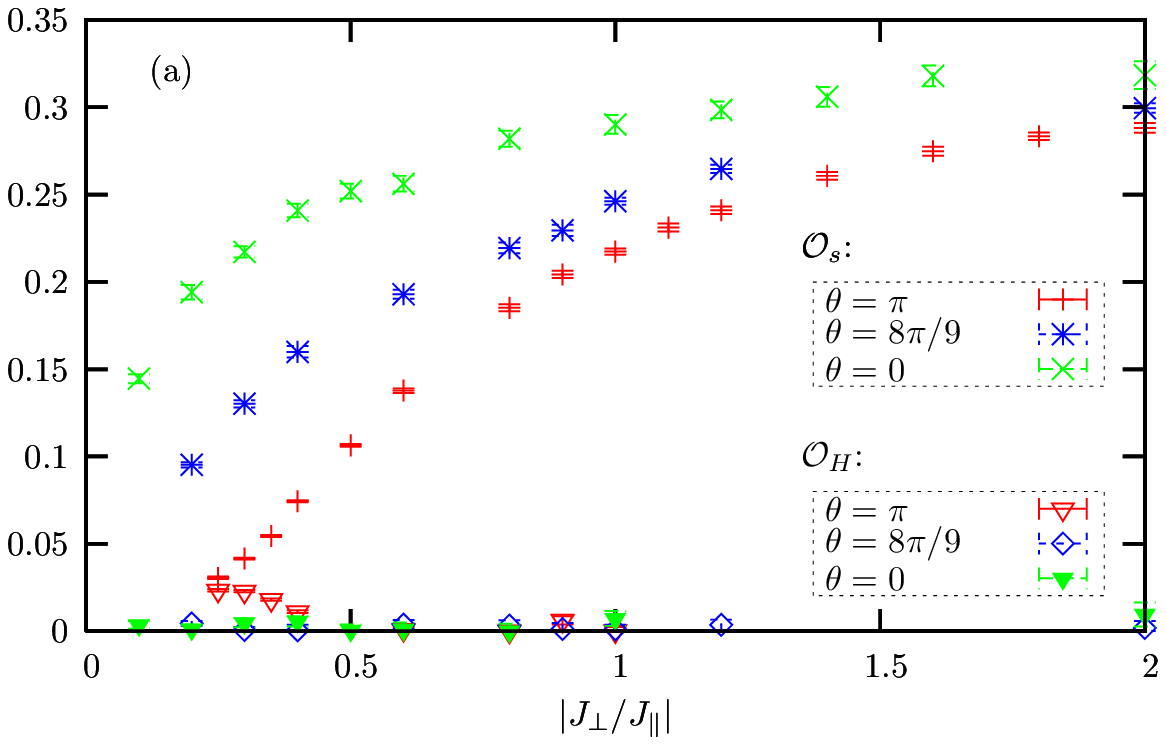}\\
\includegraphics[width=0.45\textwidth]{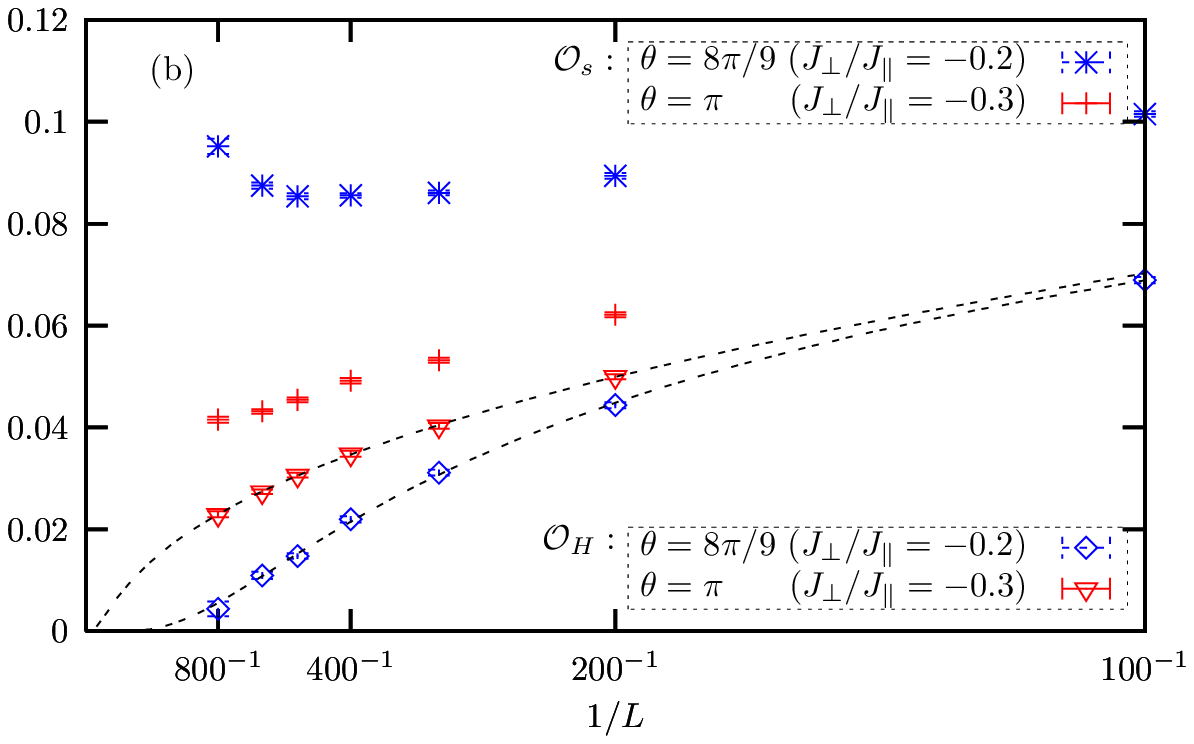}
\caption{\label{String.fig} (Color online)
         (a) String order parameter $\mathcal{O}_{s}$ and $\mathcal{O}_{H}$
     as a function  coupling $J_\perp /J_\|$ and  several twist angles.
     For $\theta=8\pi/9, \pi$ finite size effects are still present
     for the considered $L=800$ lattice in the parameter range $|J_{\perp} /J_{\|}|<0.5$.
     For $|J_{\perp} /J_{\|}| > 1.0$ the system size $L=400$ is sufficiently
         large enough to guarantee convergence.  Simulations are carried out
     up to $\beta J_{\|}=7000$.
         (b) Finite size scaling of the order parameters for the parameter
     sets $J_{\bot}/J_{\|}=-0.2$, $\theta=8\pi/9$ (blue) and
     $J_{\bot}/J_{\|}=-0.3$, $\theta=\pi$ (red). The data for
     $\mathcal{O}_{H}$ are fitted to the form:
     $\mathcal{O}_{H}\propto L^{-\alpha}\exp(-L/\xi)$.}
\end{figure}
%
The above arguments and data suggest that irrespective of the twist
angle and coupling $J_{\perp}$, the  ground state of the model
corresponds to a Haldane phase. 

%
We confirm this point of view by computing the string order parameter 
${\mathcal O}_s=\langle\hat{\mathcal{O}}_{s}(n)\rangle\vert_{n=L/2}$ on a 
$ 2\times 800$ lattice (see Fig.~\ref{String.fig}a), 
which is finite in the Haldane phase~\cite{Nijs89}. Strictly speaking, 
this is not a sufficient condition to ascertain the Haldane physics since 
we also need to show that 
${\mathcal O}_{H}=\langle\exp\left[i\pi\sum^{n_{0}+n}_{j=n_{0}}
\hat{S}^{z}_{j}\right]\rangle\vert_{n=L/2}$ vanishes in the thermodynamic 
limit (when both ${\cal O }_{s}> 0 $ and ${\cal O }_{H} > 0$, 
an Ising  order is present~\cite{Nijs89}).

In the region where the correlation length $\xi$ exceeds the 
lattice length, finite-size effects are present (see caption of Fig.
\ref{String.fig}). In particular  when the lattice size is smaller
than the correlation length, both $\mathcal{O}_{H}$ and $\mathcal{O}_{s}$ take non-zero values, since 
the very slow decay of the spin correlations  mimics Ising type order. As the system size grows beyond
the correlation length, $\mathcal{O}_{H}$ decreases exponentially 
whereas $\mathcal{O}_{s}$  in enhanced. Those size effects  are
explicitly shown in Fig. \ref{String.fig}b  at $J_\perp /J_\| =
-0.2$, $\theta = 8\pi/9$  where $L \gg \xi $ and $J_\perp /J_\| =
-0.3$, $\theta = \pi$  where our maximal system size barely exceeds
the estimated correlation length. Taking those size effects into
account, we conclude that in the thermodynamic limit, only the string order parameter
$\mathcal{O}_{s}$ is finite  in the whole $(\theta, J_\perp)$
plane.

%
%
%

In conclusion  we have established that the ferromagnetic spiral
staircase is a Haldane system, irrespective on the twist $\theta$
and coupling constant $J_\bot$.  In the weak coupling region,
twisting the ladder introduces a new low energy scale which we
interpret in terms of a SN interaction.  As a consequence and
for $\theta > \theta_c \sim  8 \pi/9$,  we have provided
numerical data showing that at weak coupling, the spin gap
decreases quicker than the linear $J_\perp$ behavior of the 2-leg ladder ($\theta=0$).  
Analysis of the data is consistent with the
picture that, for $\theta \geq \theta_c$,  the spin gap tracks the SN
scale and is hence proportional to $J_\perp^2/J_\|$.

%
%
%
We are grateful to K. Beach, K. Kikoin, P. Pujol and S. Kehrein for numerous
fruitful discussions. We are thankful to F. Essler for providing us
additional details regarding the publication~\cite{Essler07}. The
continuous time QMC simulations were performed using the looper
code~\cite{Todo01} (see {\tt
http://wistaria.comp-phys.org/alps-looper}) from the ALPS
libraries~\cite{Alet05} (see {\tt
http://alps.comp-phys.org}). We thank IDRIS (Orsay), 
CALMIP (Toulouse) and LRZ-M\"unich for use of supercomputer facilities. CB 
acknowledges financial support from the DFG under the grant
number AS120/4-2. SC and FA are supported by the French ANR
program. MNK appreciates support from the Heisenberg program of
the DFG and the SFB-410 research grant and  acknowledges support
from U.S. DOE, Office of Science, under Contract No.
DE-AC02-06CH11357. DNA thanks ICTP for the hospitality.
SC and FFA benefit from a European exchange program (Procope). 
%
%


\begin{thebibliography}{10}
\bibitem{com1} Although there is no Kondo physics involved in the
model (\ref{hh1}), we preserve the historical name "Kondo necklace" 
given by Doniach \cite{Don77}.
\bibitem{com4} 
We note that  when increasing the twist for $\theta=0$ to $\theta = \pi/2$  at  
values of $|J_{\perp}| /J_{\|} < 1 $ the spin gap 
grows (see Fig. \ref{Gap.fig}).  This  could be  consistent with  logarithmic  
corrections to the gap which  depend on the twist.  
It is beyond the scope of this numerical study to  confirm this.
\bibitem{com2}
By use of the fermionic theory \cite{kak05a}, an estimate for the critical angle
$\pi - \theta_c \sim
|J_\perp/J_\parallel|$ can be found for the case $|J_\perp|\ll
J_\parallel$  as the
angle corresponding to a diverging mass
at the $\Gamma$-point of the 1D Brillouin zone. For
$\theta>\theta_c$, the spectrum curvature is negative while for
$\theta< \theta_c$ it is positive. Therefore, the corresponding
critical exchange coupling along the second leg $J_\parallel
\cos^2(\theta_c/2)$ scales as $J_\perp^2/J_\parallel$. The details
of the analysis will be given elsewhere \cite{Brunger07}.
\bibitem{com3} Long range interaction along the second leg
is referred as RKKY (Ruderman-Kittel-Kasuya-Yosida) interaction in
the Ref. \onlinecite{Essler07}. Reserving the notation of RKKY for the
models of itinerant electrons mediating the spin-spin interaction,
we use the terminology of Suhl-Nakamura interaction for the models
of interacting local spins as more appropriate.
\end{thebibliography}
%
\end{document}